\documentclass[conference]{IEEEtran}
\usepackage[T1]{fontenc}
\usepackage[utf8]{inputenc}
\usepackage[binary-units=true]{siunitx}
\DeclareSIUnit{\bits}{bits}
\usepackage{amssymb}
\usepackage[pdftex]{graphicx}
\usepackage{subcaption} 
\usepackage{amsmath}
\usepackage{amsthm}
\usepackage{mathtools}
\usepackage{array}
\usepackage[noadjust]{cite}
\usepackage{tikz}
\usepackage[bookmarks=false]{hyperref}
\usepackage[acronym]{glossaries}
\usepackage{hhline}
\usepackage{multirow}
\usepackage{siunitx}
\usepackage{pgfplots}
\usepackage[ruled,vlined]{algorithm2e}

\newacronym{3GPP}{3GPP}{3rd Generation Partnership Project}
\newacronym{ACM}{ACM}{adaptive coding and modulation}
\newacronym{ACLR}{ACLR}{adjacent channel leakage ratio}
\newacronym{ADC}{ADC}{analog-to-digital conversion}
\newacronym{AGC}{AGC}{automatic gain control}
\newacronym{AWGN}{AWGN}{additive white Gaussian noise}
\newacronym{BER}{BER}{bit error rate}
\newacronym{BS}{BS}{base station}
\newacronym{BLER}{BLER}{block error rate}
\newacronym{BCE}{BCE}{binary cross-entropy}
\newacronym{BICM}{BICM}{bit-interleaved coded modulation}
\newacronym{BMD}{BMD}{bit-metric decoding}
\newacronym{BP}{BP}{backpropagation}
\newacronym{BPTT}{BPTT}{backpropagation through time}
\newacronym{CE}{CE}{cross-entropy}
\newacronym{CFO}{CFO}{carrier frequency offset}
\newacronym{CNN}{CNN}{convolutional neural network}
\newacronym{CCDF}{CCDF}{complementary cumulative distribution function}
\newacronym{CSI}{CSI}{channel state information}
\newacronym{DAC}{DAC}{digital-to-analog conversion}
\newacronym{DPD}{DPD}{digital pre-distortion}
\newacronym{DL}{DL}{deep learning}
\newacronym{DFT}{DFT}{discrete Fourier transform}
\newacronym{ELU}{ELU}{exponential linear unit}
\newacronym{FFT}{FFT}{fast Fourier transform}
\newacronym{GAN}{GAN}{generative adversarial network}
\newacronym{GRU}{GRU}{gated recurrent unit}
\newacronym{iid}{i.i.d.\@}{independent and identically distributed}
\newacronym{IFFT}{IFFT}{inverse fast Fourier transform}
\newacronym{KL}{KL}{Kullback-Leibler}
\newacronym{LLR}{LLR}{log-likelihood ratio}
\newacronym{LSTM}{LSTM}{long short-term memory}
\newacronym{LDPC}{LDPC}{low-density parity-check}
\newacronym{LMMSE}{LMMSE}{linear minimum mean squared error}
\newacronym{MDP}{MDP}{Markov decision process}
\newacronym{ML}{ML}{machine learning}
\newacronym{MLP}{MLP}{multilayer perceptron}
\newacronym{MIMO}{MIMO}{multiple-input multiple-output}
\newacronym{MU-MIMO}{MU-MIMO}{multi-user multiple-input multiple-output}
\newacronym{MU}{MU}{multi-user}
\newacronym{MSE}{MSE}{mean squared error}
\newacronym{NN}{NN}{neural network}
\newacronym{NR}{NR}{new radio}
\newacronym{NLOS}{NLOS}{non-line of sight}
\newacronym{OFDM}{OFDM}{orthogonal frequency-division multiplexing}
\newacronym{pdf}{pdf}{probability density function}
\newacronym{pmf}{pmf}{probability mass function}
\newacronym{PA}{PA}{power amplifier}
\newacronym{PAPR}{PAPR}{peak-to-average power ratio}
\newacronym{PSD}{PSD}{power spectral density}
\newacronym{PRT}{PRT}{peak reduction tone}
\newacronym{QPSK}{QPSK}{quadrature phase-shift keying}
\newacronym{QAM}{QAM}{quadrature amplitude modulation}
\newacronym{PSNR}{PSNR}{Peak Signal to Noise Ratio}
\newacronym{RBF}{RBF}{Rayleigh block-fading}
\newacronym{RB}{RB}{resource block}
\newacronym{RE}{RE}{resource element}
\newacronym{RG}{RG}{resource grid\newacronym{RE}{RE}{resource element}}
\newacronym{ReLU}{ReLU}{rectified linear unit}
\newacronym{RTN}{RTN}{radio transformer network}
\newacronym{RL}{RL}{reinforcement learning}
\newacronym{RNN}{RNN}{recurrent neural network}
\newacronym{SFO}{SFO}{sampling frequency offset}
\newacronym{SER}{SER}{symbol error rate}
\newacronym{SNR}{SNR}{signal-to-noise ratio}
\newacronym{SINR}{SINR}{signal-to-interference-plus-noise ratio}
\newacronym{SGD}{SGD}{stochastic gradient descent}
\newacronym{SISO}{SISO}{single-input single-output}
\newacronym{SIMO}{SIMO}{single-input multiple-output}
\newacronym{SU}{SU}{single-user}
\newacronym{TDD}{TDD}{time-division duplexing}
\newacronym{TR}{TR}{tone reservation}
\newacronym{UE}{UE}{user equipment}
\newacronym{UMi}{UMi}{urban microcell}
\newacronym{wrt}{w.r.t.\@}{with respect to}

\renewcommand{\vec}[1]{\mathbf{#1}}
\newcommand{\vecs}[1]{\boldsymbol{#1}}

\newcommand{\bv}{\vec{b}}
\newcommand{\cv}{\vec{c}}

\newcommand{\xv}{\vec{x}}
\newcommand{\yv}{\vec{y}}
\newcommand{\zv}{\vec{z}}

\newcommand{\thetav}{\vecs{\theta}}
\newcommand{\psiv}{\vecs{\psi}}

\newcommand{\Bm}{\vec{B}}

\newcommand{\Fm}{\vec{F}}

\newcommand{\Vm}{\vec{V}}
\newcommand{\Wm}{\vec{W}}

\newcommand{\CC}{\mathbb{C}}

\newcommand{\RR}{\mathbb{R}}

\newcommand{\tp}{^{\mathsf{T}}}

\newcommand{\EE}{{\mathbb{E}}}

\newlength{\dhatheight}

\IEEEoverridecommandlockouts
\begin{document}
\begin{NoHyper}
\title{End-to-End Learning of OFDM Waveforms with PAPR and ACLR Constraints}

\author{
\IEEEauthorblockN{Mathieu Goutay\IEEEauthorrefmark{1}\IEEEauthorrefmark{3}, Fayçal Ait Aoudia\IEEEauthorrefmark{1}, Jakob Hoydis\IEEEauthorrefmark{2}\thanks{Work carried out while J. Hoydis was with Nokia Bell Labs.}, and Jean-Marie Gorce\IEEEauthorrefmark{3}}
\IEEEauthorblockA{\IEEEauthorrefmark{1}Nokia Bell Labs, Paris-Saclay, 91620 Nozay, France}
\IEEEauthorblockA{\IEEEauthorrefmark{2}NVIDIA, 06906 Sophia Antipolis, France}
\IEEEauthorblockA{\IEEEauthorrefmark{3}Université de Lyon, INSA Lyon, Inria, CITI,  69100 Villeurbanne, France \\
\{mathieu.goutay, faycal.ait\_aoudia\}@nokia.com, jhoydis@nvidia.com, jean-marie.gorce@insa-lyon.fr
}}

\maketitle

\begin{abstract}

\Gls{OFDM} is widely used in modern wireless networks thanks to its efficient handling of multipath environment.
However, it suffers from a poor \gls{PAPR} which requires a large power backoff, degrading the \gls{PA} efficiency.
In this work, we propose to use a \gls{NN} at the transmitter to learn a high-dimensional modulation scheme allowing to control the \gls{PAPR} and \gls{ACLR}. On the receiver side, a \gls{NN}-based receiver is implemented to carry out demapping of the transmitted bits.
The two \glspl{NN} operate on top of \gls{OFDM}, and are jointly optimized in and end-to-end manner using a training algorithm that enforces constraints on the \gls{PAPR} and \gls{ACLR}.
Simulation results show that the learned waveforms enable higher information rates than a tone reservation baseline, while satisfying predefined \gls{PAPR} and \gls{ACLR} targets.
\end{abstract}

\glsresetall

\section{Introduction} 
\label{sec:introduction}

With the ever-growing demand for services that depend on radio connectivity, future wireless systems will need to satisfy increasingly difficult requirements on the signal characteristics.
Due to its low complexity implementation, \gls{OFDM} is used in many modern communication systems and is a potential candidate waveform for the next generation of wireless networks.
However, when used in conjunction with conventional \gls{QAM}, the generated signal exhibits both low spectral containment and high \gls{PAPR}.
%
The \gls{ACLR} of \gls{OFDM} waveforms is typically decreased by the introduction of guard subcarriers at the cost of a reduced spectral efficiency.
High \glspl{PAPR} either leads to distortions of the transmitted signal or to the use of high power backoffs, reducing the \gls{PA} efficiency.

Multiple techniques have been proposed to reduce the \gls{PAPR} of \gls{OFDM} signals, such as clipping and filtering \cite{wang2010optimized}, constellation extension \cite{krongold2003reduction}, and tone reservation \cite{1261335}.
The latter approach uses a subset of the available subcarriers to create a peak-canceling signal, and was recently combined with deep \glspl{NN} to predict the peak-canceling signal from the frequency baseband symbols to be transmitted~\mbox{\cite{wang2020novel, wang2021model, 8928103}}.
In~\cite{8240644}, it has been proposed to treat the communication system as an autoencoder, where the transmitter and receiver are trained to jointly minimize the \gls{BER} and \gls{PAPR} of the transmission.
Although the presented autoencoder shows promising results, the \gls{PAPR} is minimized based on a non-oversampled time-discrete signal, which is not fully representative of its  analog waveform \cite{1261335}.
Moreover, the autoencoder operates on symbols, meaning that \gls{QAM} mapping and demapping are still required.
Finally, none of these previous works allow to define target values for the PAPR and ACLR, and therefore to control the tradeoff between the spectral efficiency and the PAPR and ACLR requirements.


In this work, we propose to jointly optimize an \gls{NN}-based high-dimensional modulation scheme on the transmitter side and an NN-based detector on the receiver side, that operate on top of OFDM. 
The end-to-end system is trained to maximize an achievable information rate, while satisfying constraints on the ACLR and PAPR. 
The key idea is to formulate the training objectives as a constrained optimization problem, which is then solved iteratively using the augmented Lagrangian method.
To that aim, the achievable rate, \gls{ACLR}, and \gls{PAPR} of the transmission are expressed as functions that can be evaluated during training and minimized using \gls{SGD}.
%
Evaluations show that the end-to-end system achieves competitive or higher rates compared to a \gls{TR} baseline, while having similar or lower ACLRs and PAPRs.
As an example, the trained system outperforms the \gls{TR} baseline while having a 0.7~\si{dB} and 10~\si{dB} reduced PAPR and ACLR, respectively.
%


\section{Problem positioning} 
\label{sec:problem_positioning}

\subsection{System model}
\label{sec:system_model}

An \gls{OFDM} system with $N$ subcarriers is considered, the subcarriers being indexed by the set $\mathcal{N}= \{ -\frac{N-1}{2}, \dots, \frac{N-1}{2} \}$, where $N$ is assumed to be odd for convenience.
The matrix of bits to be transmitted is denoted by $\Bm = \left[ \bv_1, \dots, \bv_N \right]\tp$, where $\bv_{n\in\mathcal{N}} \in \{0, 1\}^{K}$ are the vectors of bits to be sent and $K$ is the number of bits per channel uses.
$\Bm$ is modulated onto discrete baseband symbols $\xv \in \CC^{N}$ that are mapped on the orthogonal subcarriers, forming the baseband spectrum
\begin{equation}
\begin{split}
S(f) & = \sum_{n=-\frac{N-1}{2}}^{\frac{N-1}{2}} x_n \frac{1}{\sqrt{\Delta_f}} \text{sinc} \left( \frac{f}{\Delta_f} - n \right) . \\
\end{split}
\end{equation}
where $\Delta_f$ is the subcarrier spacing.
The corresponding time-domain signal, of duration $T=\frac{1}{\Delta_f}$, is
\begin{equation}
\begin{split}
s(t) & = \sum_{n=-\frac{N-1}{2}}^{\frac{N-1}{2}} x_n \frac{1}{\sqrt{T}} \text{rect} \left( \frac{t}{T} \right) e^{i 2 \pi n t/T}  \\
\end{split}
\end{equation}
where $t \in [-\frac{T}{2}, \frac{T}{2}]$. 

A major limitation of \gls{OFDM} is its high PAPR, which either implies that the output signal might suffer from distortions due to the \gls{PA} saturation, or that the \gls{PA} must be operated with a large power back-off, reducing its power efficiency \footnote{A  substantial power backoff is required although \gls{DPD} is widely used to reduce the detrimental effect of such distortions.}.
In the following, we define the \gls{PAPR} of a signal as the smallest threshold $e \geq 0$, such that the probability of the ratio between the instantaneous and average squared signal amplitude being larger than that $e$ is smaller than a threshold $\epsilon\in(0,1)$:
%
\begin{equation}
\label{eq:papr}
\begin{split}
\text{PAPR}_{\epsilon}  \coloneqq \mathrm{min} \; e, \;\; \text{s. t.} \;\;  P\left( \frac{|s(t)|^2}{\EE \left[|s(t)|^2 \right]} > e  \right) \leq \epsilon.
\end{split}
\end{equation}
%
We do not use the more conventional definition $ \frac{\max{|s(t)|^2}}{\EE \left[|s(t)|^2 \right]}$, as it only considers the max signal amplitude and is therefore less representative of the overall power distribution.

To quantify the spectral containment of waveforms, one typically uses the \gls{ACLR}, which is defined as the ratio between the expected out-of-band energy $\EE_{\xv} \left[ E_O\right]$ and the expected in-band energy $\EE_{\xv} \left[ E_I\right]$ :
\begin{equation}
\begin{split}
\text{ACLR} & \coloneqq  \frac{\EE_{\xv} \left[ E_O\right]}{\EE_{\xv} \left[ E_I\right]} 
 =   \frac{\EE_{\xv} \left[ E_A \right]}{\EE_{\xv} \left[ E_I \right]}-1 \\
\end{split}
\end{equation}
where $E_A = E_O + E_I $ is the total energy of the transmitted signal.
The in-band energy $E_I$ is
\begin{equation}
\begin{split}
E_{I} & \coloneqq \int_{-\frac{N \Delta_f}{2}}^{\frac{N \Delta_f}{2}} \left| S (f)\right|^2 df = \xv ^H \Vm \xv, \\
\end{split} 
\end{equation}
where each element $v_{a, b}$ of the matrix $\Vm$ is 
\begin{equation}
v_{a, b} = \frac{1}{\Delta_f} \int_{-\frac{N \Delta_f}{2}}^{\frac{N \Delta_f}{2}}  \text{sinc} \left(\frac{f}{\Delta_f}-a \right) \text{sinc} \left(\frac{f}{\Delta_f}-b \right) df.
\end{equation}
The total transmitted energy can be similarly computed in the time domain :
\begin{equation}
\begin{split}
E_{A} & \coloneqq \int_{ -\frac{T}{2}}^{\frac{T}{2}} \left| s (t)\right|^2 dt = \xv ^H \Wm \xv, \\
\end{split} 
\end{equation}
where $\Wm$ is such that 
\begin{equation}
w_{a, b} = \frac{1}{T} \int_{-\frac{T}{2}}^{\frac{T}{2}}  e^{i 2 \pi (a-b) t /T} dt .
\end{equation}
In the following, an \gls{AWGN} channel is considered, and the vector of received baseband symbols is denoted by $\yv \in \CC^{N}$.

\subsection{Baseline}
\label{sec:baseline}

One technique to reduce the \gls{PAPR} of \gls{OFDM} signals is tone reservation (TR), in which a subset of $R$ subcarriers (tones) are used as \glspl{PRT}.
These tones do not carry data, but are used to minimize the peak amplitude of the time-domain signal.
The set of all \glspl{PRT} is denoted by $\mathcal{R} \subset \mathcal{N}$, and the set containing the remaining $D$ subcarrier used for data transmission is denoted by $\mathcal{D} \subseteq \mathcal{N} $, with $\mathcal{D} \cup \mathcal{R} = \mathcal{N}$.
The vector of signals mapped to the \glspl{PRT} is referred to as the reduction signal $\cv\in\CC^{N}$, and $\xv$ and $\cv$ are such that $x_{n\in \mathcal{R}} = 0$ and $c_{n\in \mathcal{D}} = 0 $.
The vector of transmitted baseband symbols is denoted by $\xv' = \xv + \cv$.

\begin{figure}
\input{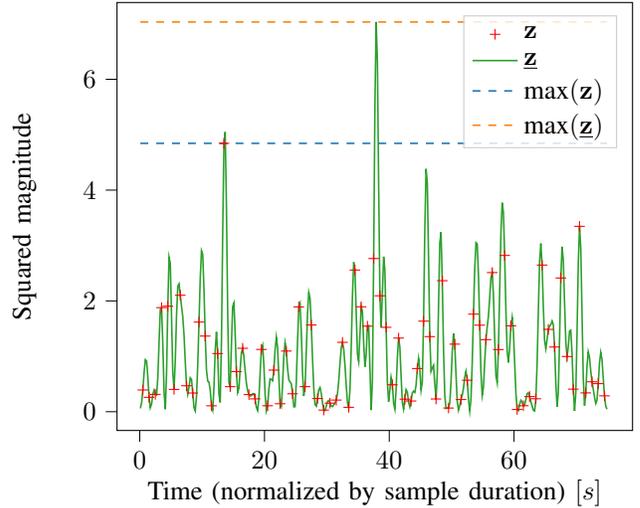}
\caption{OFDM signal generated from $N=75$ subcarriers.}
\label{fig:waveform}
\end{figure}

The corresponding continuous-time signal is denoted by $s'(t)$ and minimizing its PAPR amounts to finding 
\begin{equation}
\label{eq:c_star}
\begin{split}
\arg\min_{\cv} \max_{t} |s'(t)|^2.
\end{split}
\end{equation}

In practice, minimization cannot be performed directly on the time-continuous signal, and $s'(t)$ must therefore be discretized.
Many previous works only consider a non-oversampled discrete signal to represent the analog waveform~{\cite{wang2021model, 8928103, 8240644}}.
However, it has been shown that oversampling is necessary to correctly represent the underlying continuous signal \cite{1261335}.
Let us denote by $\zv\in\CC^N$ and $\underline{\zv}\in\CC^{NO_s}$ the signal sampled with a period $\frac{T}{N}$ and the signal oversampled by a factor $O_s$, respectively.
The difference between them is illustrated in Fig.~\ref{fig:waveform}, where the time-domain signal of an OFDM symbol is represented with and without oversampling ($O_s=5$ in this example).
First, it can be observed that the two signals have very different power peaks, indicated by dashed lines.
Second, $\underline{\zv}$ shows many more secondary peaks, which might lie in the \gls{PA} saturation region and therefore must be considered when reducing the signal distortion.
This oversampled signal can be obtained by computing the inverse discrete Fourier transform (IDFT) of $\xv'$ with adequate \mbox{zero-padding :}
\begin{equation}
\underline{\zv} = \Fm^{-1} (\xv + \cv),
\end{equation}
where $\Fm^{-1}\in\CC^{NO \times N}$ is the corresponding IDFT matrix.

The problem \eqref{eq:c_star} can be approximated by minimizing on the oversampled signal, i.e., 
\begin{equation}
\label{eq:convex_min}
  \arg\min_{\cv} \left\Vert g\left(\Fm^{-1} (\xv + \cv)\right)  \right\Vert_{\infty}
\end{equation}
where $g(\cdot)$ computes the element-wise squared magnitude $|\cdot|^2$.
Numerous iterative algorithms have been proposed to approximately solve \eqref{eq:convex_min} (see, e.g., \cite{1261335}). 
As the function \mbox{$h(\xv')= \left\Vert g\left(\Fm^{-1} (\xv')\right)  \right\Vert_{\infty}$} is convex, we use a convex solver~\cite{diamond2016cvxpy} to solve \eqref{eq:convex_min} for each vector of symbols $\xv$.
Moreover, it was shown that placing PRTs at randomly sampled locations at each transmission leads to the lowest PAPR~\mbox{\cite{peterssonPAPR, 1261335}}.
We therefore used such a random PRT allocation as baseline.
Note that neither using a convex solver nor a random PRT allocation is practical because of the incurred complexity and overhead required to indicate the \glspl{PRT} positions to the receiver.
However, it was considered for benchmarking our approach as it provides the lowest PAPR.


Aside from leveraging \gls{TR}, the baseline uses a conventional \gls{QAM} modulation to map each vector of bits $\bv_{n\in\mathcal{D}}$ to a baseband symbol $x_{n\in\mathcal{D}}$.
These symbols are then transmitted using an OFDM waveform.
Because an \gls{AWGN} channel is considered, the receiver recovers the baseband symbols $\yv$ by performing a DFT on the subcarriers $n\in{\mathcal{D}}$.
The \glspl{LLR} are then computed using a standard \gls{AWGN} demapper.

\section{Learning a high-dimensional modulations}
\label{sec:autoenoder}

\subsection{Problem formulation}

\begin{figure}[t]
    \centering
    \includegraphics[width=0.42\textwidth]{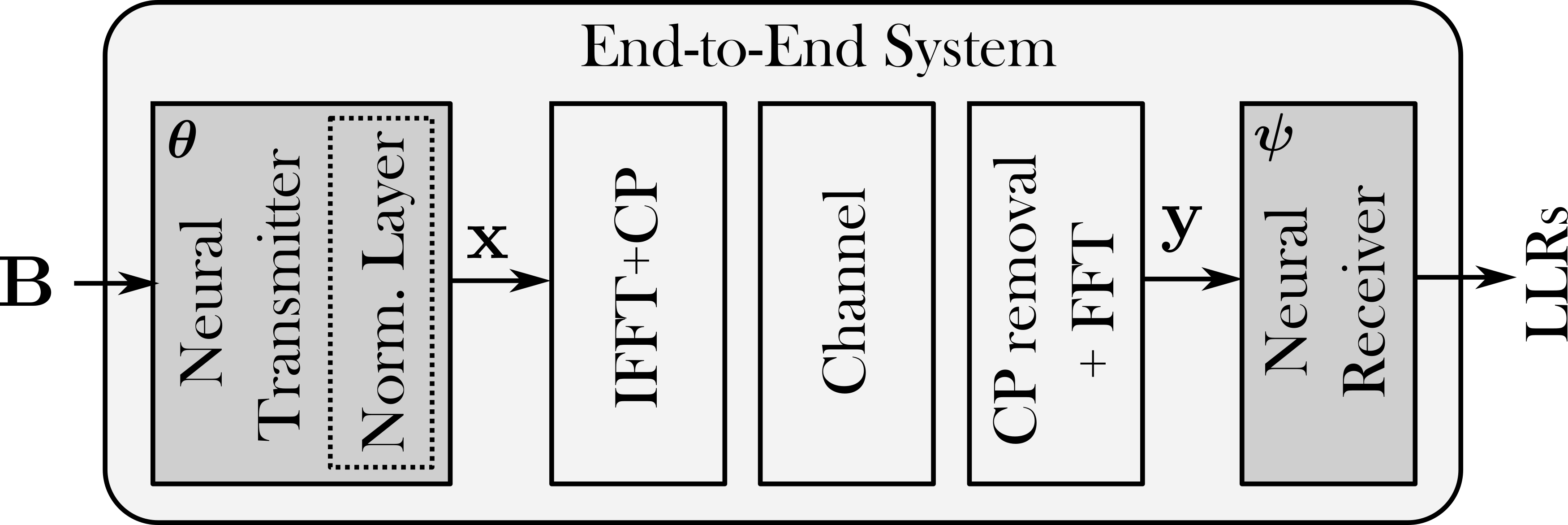}
    \caption{Trainable system, where grayed blocks represent trainable components.}
    \label{fig:system2}
\end{figure}

The proposed end-to-end system architecture  is depicted in Fig.~\ref{fig:system2}.
It is composed of a neural tranmsitter and neural receiver,  having trainable parameters respectively denoted by $\thetav$ and $\psiv$ and that operate on top of OFDM.
In the following, the end-to-end system will be referred to as "E2E" system for brevity.
We aim at optimizing the E2E system to design a high-dimensional modulation and a corresponding detector that maximize an achievable rate under \gls{PAPR} and \gls{ACLR} constraints.
The considered rate is \cite{pilotless20}
\begin{align}
\label{eq:rate0}
C(\thetav, \psiv) & = \frac{1}{N}\sum_{n=0}^{N-1}\sum_{k=0}^{K-1}I\left( b_{n,k}; \yv \right | \thetav)\\
& - \frac{1}{N}\sum_{n=0}^{N-1}\sum_{k=0}^{K-1}\EE_{\yv}\left[\text{D}_\text{KL} \left( P (b_{n,k}|\yv)|| \widehat{P}_{\psiv}(b_{n,k}| \yv ) \right) \right] \nonumber
\end{align}
which was shown to be achievable assuming a bit metric decoder \cite{bocherer2017achievable}.
The first term in \eqref{eq:rate0} is the mutual information between $b_{n,k}$ and $\yv$, and corresponds to an achievable information rate assuming an ideal receiver. 
The second term in \eqref{eq:rate0} is the KL-divergence between the true posterior probability $P (b_{n,k}|\yv)$ and the one estimated by our \gls{NN}-receiver $\widehat{P}_{\psiv}(b_{n,k}| \yv )$. 
Such probabilities can be obtained from the \gls{LLR} corresponding to the bit $b_{n,k}$, which is defined by
\begin{equation}
\text{LLR}(n,k) \coloneqq \text{ln} \left( \frac{\widehat{P}_{\psiv}(b_{n,k} = 1| \yv ) }{\widehat{P}_{\psiv}(b_{n,k} = 0| \yv ) } \right).
\end{equation}
Intuitively, the KL-divergence term corresponds in \eqref{eq:rate0} to a rate loss due to an imperfect receiver.

Let us denote by $\gamma_{\text{peak}}$ the targeted PAPR and by $\beta_{\text{leak}}$ the targeted ACLR.
In the E2E system, both quantities depend on the trainable parameters of the neural transmitter.
Maximizing the rate under these constraints can be formulated as the following optimization problem :
\begin{subequations}
\label{eq:rate}
 \begin{align}
\underset{\thetav, \psiv}{\text{maximize}} & \quad\quad C(\thetav, \psiv) \label{eq:rate1} \\
\text{subject to} & \quad\quad \frac{1}{N}  \EE_{\xv}[E_A] = 1  \label{eq:rate2}\\
& \quad\quad \text{PAPR}_{\epsilon}(\thetav) = \gamma_{\text{peak}} \label{eq:rate3} \\
&  \quad\quad \text{ACLR}(\thetav) \leq \beta_{\text{leak}} \label{eq:rate4} 
 \end{align}
\end{subequations}
where \eqref{eq:rate2} ensures that the average energy per OFDM symbol equals $N$.

\subsection{System training}

Solving the problem \eqref{eq:rate} is achieved by using the augmented Lagrangian method detailed in \cite[Chapter~3]{bertsekas2014constrained}.
This method converts the constrained optimization problem into its augmented Lagrangian that is minimized with respect to the trainable parameters $\thetav$ and $\psiv$.
Because the E2E system is composed of an \gls{NN}-based transmitter and receiver, their parameters can be optimized using \gls{SGD} as long as the loss function is differentiable. 
In this section, we therefore express the objective \eqref{eq:rate1} and the constraints \eqref{eq:rate3} and \eqref{eq:rate4} as differentiable functions that can be evaluated during training and minimized with \gls{SGD}.
The constraint \eqref{eq:rate2} is enforced with a normalization layer that is presented in the following.

To ensure a unitary mean energy per symbol, such normalization layer is added to the neural transmitter (see Fig.\ref{fig:system2}) and need to perform
\begin{align}
\label{eq:layer_norm*}
l_{\text{norm}}^*(\xv) = \frac{\xv}{ \left( \frac{1}{N} \EE_{\xv}[E_A] \right)^{\frac{1}{2}} } .
\end{align}
Computing the true value of $\EE_{\xv}[E_A]$ is of prohibitive complexity due to the $2^{KN}$ combinations of bits corresponding to different vectors $\xv$.
We therefore normalize each batch of transmitted signals to ensure it has an average energy of one. 
Each element $j$ of a batch is denoted by the superscript $[j]$ and is normalized as follows
\begin{align}
\label{eq:layer_norm}
l_{\text{norm}}(\xv^{[j]}) = \frac{\xv^{[j]}}{  \left( \frac{1}{N B_s}  \sum_{i=1}^{B_s}  \xv^{[i]^{\mathsf{H}}} \Wm \xv^{[i]} \right)^{\frac{1}{2}} }
\end{align}
%
This expression accounts for the correlation that can appear  between the frequency baseband symbols generated by the neural transmitter.
Such correlation would not occur in conventional bit-interleaved coded modulation systems, as the baseband symbols would be independent and identically distributed.

As derived in \cite{9118963}, maximizing the rate  of the communication system \eqref{eq:rate1} is equivalent to minimizing the total \gls{BCE}
\begin{align}
\label{eq:CE}
L_C(\thetav, \psiv) &:= - \frac{1}{N}\sum_{n=0}^{N-1}\sum_{k=0}^{K-1} \EE_{\yv} \left[ \text{log}_2 \left(\widehat{P}_{\psiv} (b_{n,k}| \yv ) \right) \right]  \\
& = K -  C(\thetav, \psiv). \nonumber
\end{align}
Because the exact computation of the \gls{BCE} value would be of prohibitive complexity, a common practice is to estimate it using Monte Carlo sampling with batches of size $B_s$:
\begin{align}
\label{eq:CE_batch}
L_C(\thetav, \psiv) \approx & - \frac{1}{NB_s}\sum_{n=0}^{N-1}\sum_{k=0}^{K-1}\sum_{i=0}^{B_s-1}  \text{log}_2 \left( \widehat{P}_{\psiv} \left( b_{n,k}^{[i]}| \yv ^{[i]} \right) \right).
\end{align}
%

The PAPR constraint \eqref{eq:rate3} can be interpreted as producing time-domain signals whose power $|s(t)|^2$ exceeds $\gamma_{\text{peak}}$ with a low probability.
However, evaluating \eqref{eq:papr} for arbitrary values of $\epsilon$ requires to count the amount of samples whose energy are higher than $\gamma_{\text{peak}}$, which is not differentiable.
We propose to substitute \eqref{eq:rate3} by a constraint function that penalizes all signals exceeding the threshold $\gamma_{\text{peak}}$, which is equivalent to setting $\epsilon=0$ :
\begin{align}
\label{eq:loss_papr}
L_{\gamma_{\text{peak}}}(\thetav)  = \EE \left[ \int_{-\frac{T}{2}}^{\frac{T}{2}} \text{max}\left(0, |s(t)|^2-\gamma_{\text{peak}} \right) dt \right].
\end{align}
Similarly to Section~\ref{sec:baseline}, let us denote by $\underline{\zv}  = \Fm^{-1}\xv \; \in \CC^{NO_s}$ the vector of discrete time signal oversampled with a factor $O_s$, where $\xv\in\CC^{N}$ is the output of the neural transmitter.
The expectation in \eqref{eq:loss_papr} can be approximated by sending a batch of signals, while the integral can be approximated using a Riemann sum, leading to
\begin{align}
L_{\gamma_{\text{peak}}}(\thetav)  \approx \frac{T}{B_s N O} \sum_{i=0}^{B_s-1} \sum_{t=-\frac{NO_s-1}{2}}^{\frac{NO_s-1}{2}} \text{max} \left(0, \left| \underline{z}_t^{[i]} \right| ^2 - \gamma_{\text{peak}} \right) .
\end{align}

Finally, inequality constraints such as \eqref{eq:rate4} can be converted to equality constraints using slack variables
$q\in\RR_+$:
\begin{align}
\text{ACLR} \leq \beta_{\text{leak}} \iff \text{ACLR} - \beta_{\text{leak}}  = -q .
\end{align}
This constraint can then be enforced my minimizing $L_{\beta_{\text{leak}}}(\thetav) + q $, with
\begin{align}
L_{\beta_{\text{leak}}}(\thetav)  & =  \frac{ \EE \left[ E_A \right]}{ \EE \left[ E_I \right]}-1   - \beta_{\text{leak}} \\
& \approx \frac{  \frac{1}{B_s} \sum_{i=0}^{B_s-1}  \xv^{[i]^{\mathsf{H}}} \Wm \xv^{[i]}}{ \frac{1}{B_s} \sum_{i=0}^{B_s-1}  \xv^{[i]^{\mathsf{H}}} \Vm \xv^{[i]}} -1   - \beta_{\text{leak}} .
\end{align}

The problem \eqref{eq:rate} can then be reformulated for $\epsilon = 0$ as the following constrained optimization problem:
\begin{subequations}
\label{eq:pb}
 \begin{align}
\underset{\thetav, \psiv}{\text{minimize}} & \quad\quad L_C(\thetav, \psiv) \label{eq:pb1} \\
\text{subject to} & \quad\quad L_{\gamma_{\text{peak}}}(\thetav)  = 0 \label{eq:pb2} \\
&  \quad\quad L_{\beta_{\text{leak}}}(\thetav) + q \leq 0 \label{eq:pb3} 
 \end{align}
\end{subequations}
for which the augmented Lagrangian is \cite[Chapter~3]{bertsekas2014constrained}
%
\begin{align}
\label{eq:lagrange}
\overline{L} (\thetav, \psiv, \lambda_p, \lambda_l, & \mu_p, \mu_l)  = L_C(\thetav, \psiv) \nonumber \\
& + \lambda_p L_{\gamma_{\text{peak}}}(\thetav) + \frac{1}{2} \mu_p |L_{\gamma_{\text{peak}}}(\thetav)|^2 \\
& + \frac{1}{2\mu_l} \left( \text{max}(0, \lambda_l + \mu_l L_{\beta_{\text{leak}}}(\thetav) )^2 - \lambda_l^2 \right).  \nonumber
\end{align}
One can see that $q$ is not present in \eqref{eq:lagrange} since minimizing the augmented Lagrangian with respect to $q$ can be carried out explicitly for each fixed pair of $\{ \thetav, \psiv \}$ \cite{bertsekas2014constrained}.
The quantities $\mu_p > 0$ and $\mu_l>0$ are the penalty parameters and $\lambda_p, \lambda_l$ are the Lagrange multipliers for the constraint functions $L_{\gamma_{\text{peak}}}(\thetav)$ and $L_{\beta_{\text{leak}}}(\thetav)$ respectively.
%
%
The augmented Lagrangian method consists in iteratively minimizing \eqref{eq:lagrange}, each iteration comprising multiple steps of \gls{SGD}  followed by an update of the hyperparameters  $\lambda_p, \lambda_l, \mu_p$, and $\mu_l$.
The training procedure is described in Algorithm 1, where $\lambda_p^{(u)}, \lambda_l^{(u)}$, $\mu_p^{(u)}$, and $\mu_l^{(u)}$ denote the value of the hyperparameters at the $u^{\text{th}}$ iteration, and $\tau \in \RR_+$ controls the evolution of $\mu_p$ and $\mu_l$.
\begin{algorithm}
\label{algo}
\SetAlgoLined
 Initialize $\thetav, \psiv, \lambda_p^{(0)}, \lambda_l^{(0)}, \mu_p^{(0)}, \mu_l^{(0)}$ \\
 \For{$u = 0, ... $}{
  $\triangleright$ Perform multiple steps of SGD \\
  on $\overline{L} (\thetav, \psiv, \lambda, \lambda_l, \mu_p, \mu_l)$ w.r.t. $\thetav$ and $ \psiv$ \\
  $\triangleright$ Update optimization hyperparameters : \\
  $\lambda_p^{(u+1)} = \lambda_p^{(u)} + \mu_p^{(u)} L_{\gamma_{\text{peak}}}(\thetav) $\\
  $\lambda_l^{(u+1)} = \text{max} \left(0, \lambda_l^{(u)}  + \mu_l^{(u)} L_{\beta_{\text{leak}}}(\thetav) \right)$\\
  $\mu_p^{(u+1)} = (1+\tau) \mu_p^{(u)}$ \\
  $\mu_l^{(u+1)} = (1+\tau) \mu_l^{(u)}$ 
 }
 \caption{Training procedure}
\end{algorithm}


\subsection{System architecture}


\begin{figure}
\centering
\begin{subfigure}{.16\textwidth}
  \centering
  \includegraphics[height=125pt]{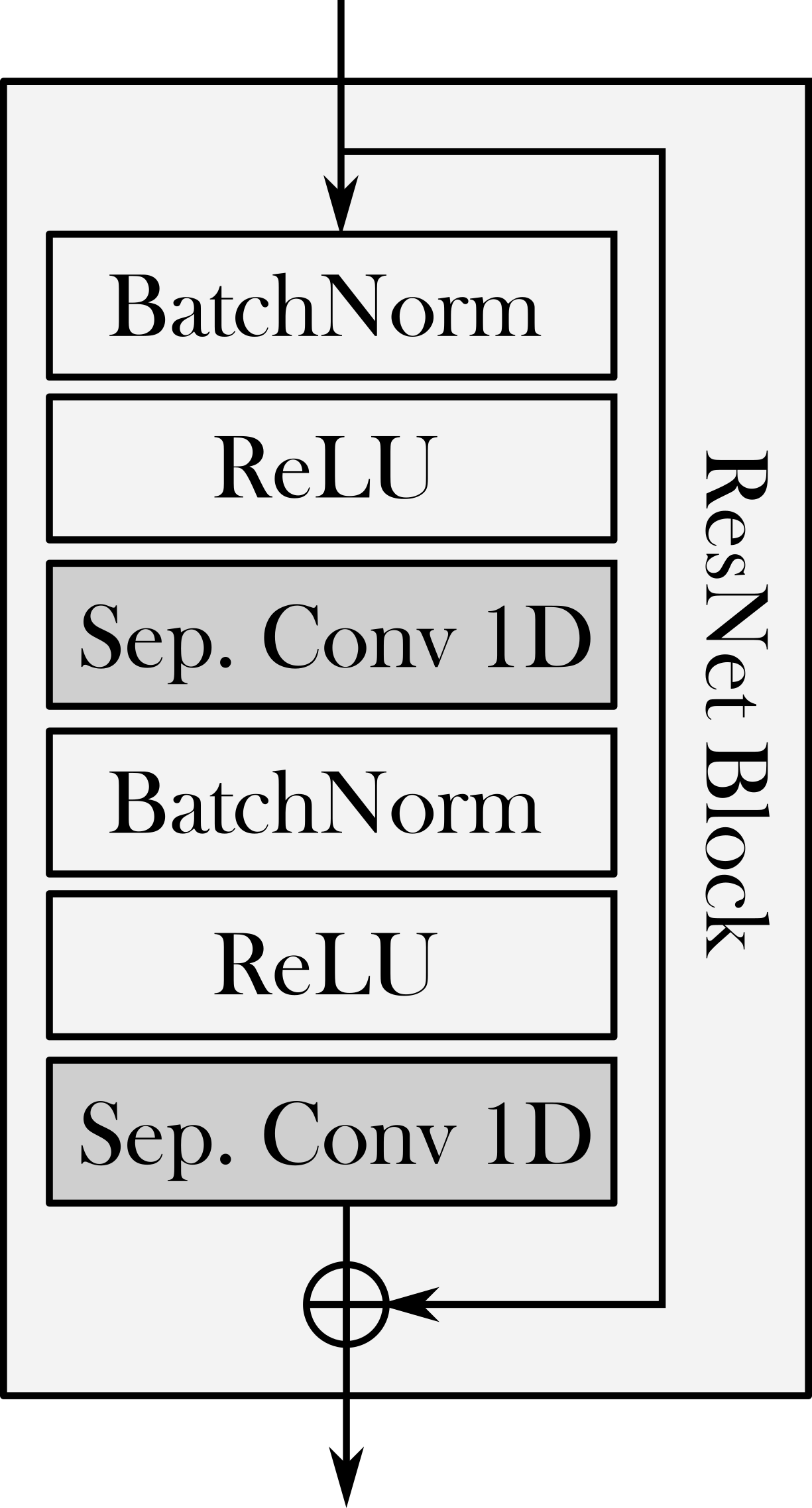}
  \caption{ResNet block.}
  \label{fig:resnet_h}
\end{subfigure}%
\begin{subfigure}{.16\textwidth}
  \centering
  \includegraphics[height=125pt]{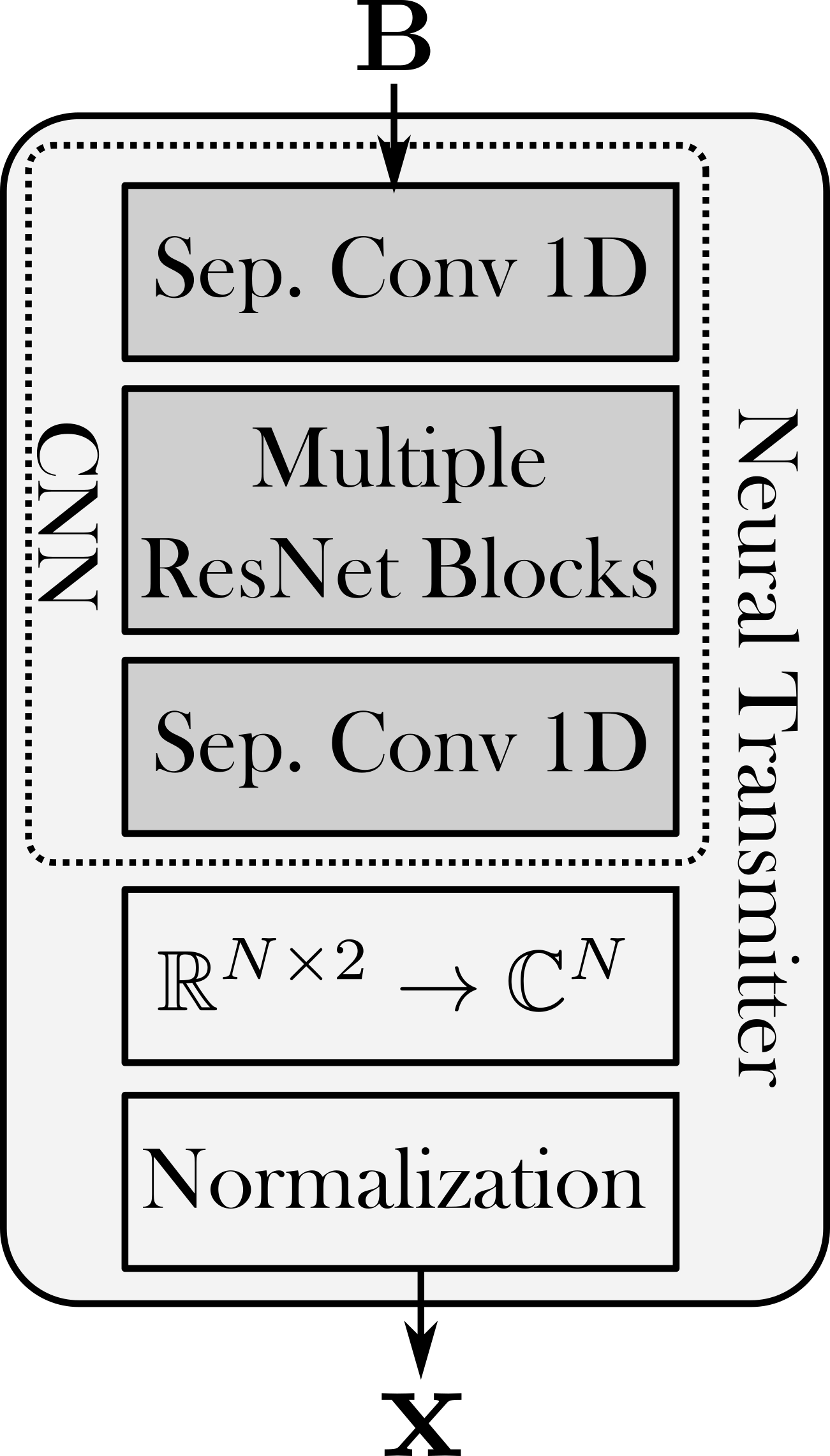}
  \caption{NN-Transmiter.}
  \label{fig:nn_tx_h}
\end{subfigure}
\begin{subfigure}{.16\textwidth}
  \centering
  \includegraphics[height=125pt]{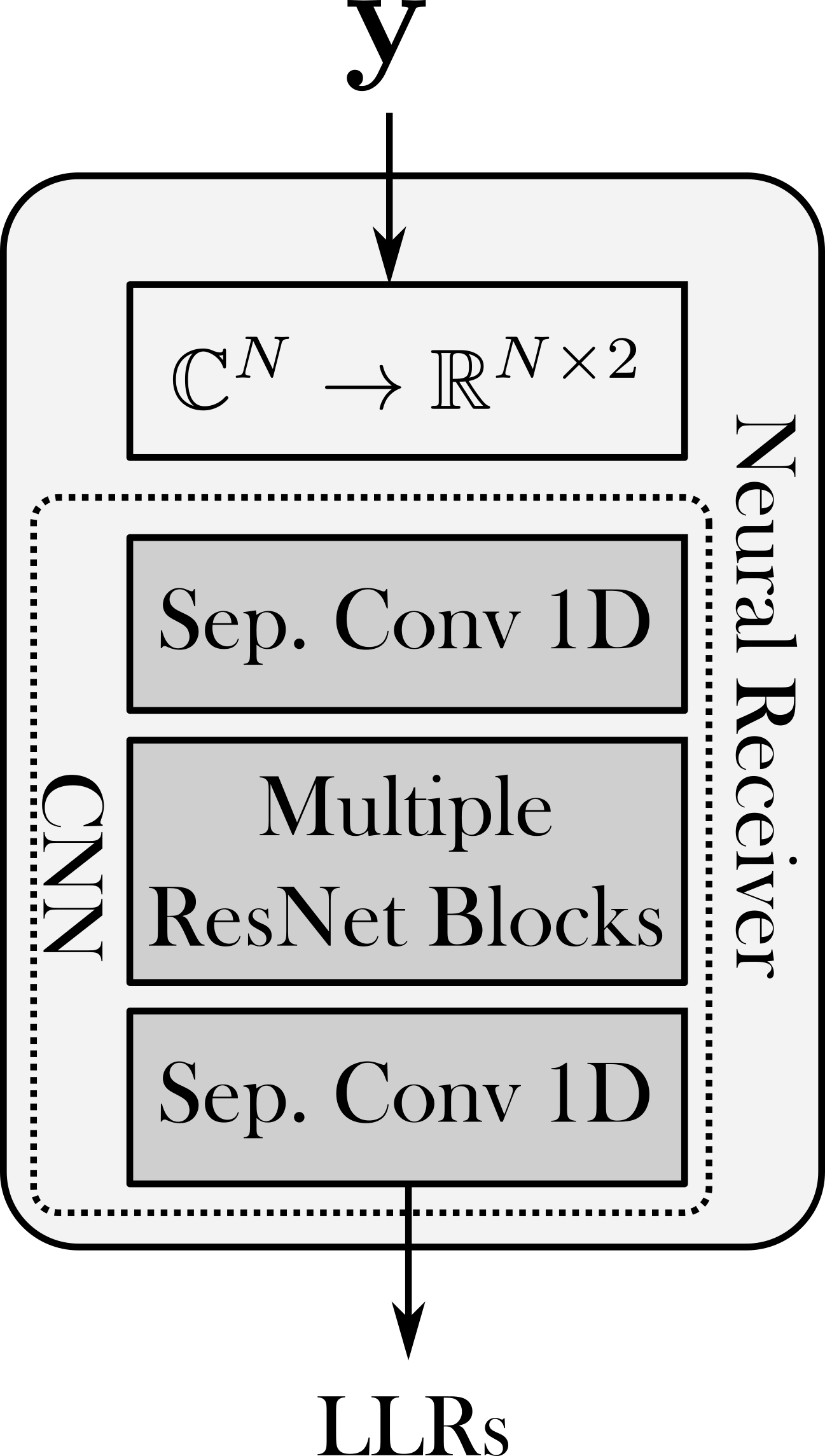}
  \caption{NN-Receiver.}
  \label{fig:nn_rx_h}
\end{subfigure}%
\caption{Different parts of the end-to-end system, where grayed blocks are trainable components. The IDFT/DFT operations and the channel are not represented for clarity.}
\label{fig:components}
\end{figure}

Both the \gls{NN}-based transmitter and receiver use ResNet blocks, originally presented in \cite{he2016identity}, and which were proven to be effective in physical layer tasks \cite{honkala2020deeprx, pilotless20, 9460758}.
A ResNet block is composed of a residual connection preceded by two groups of layers, each containing a batch normalization layer, a ReLU, and a 1D separable convolution (Fig. \ref{fig:resnet_h}). 
Such a convolution maintains similar performance compared to a traditional convolution while being less computationally demanding \cite{howard2017mobilenets}.
All 1D separable convolutions use zero-padding to maintain the first dimension of constant size.

The neural transmitter (Fig. \ref{fig:nn_tx_h}) is composed of two convolutional layers, multiple ResNet blocks, a layer that converts $2N$ real numbers to $N$ complex numbers, and a normalization layer as expressed in \eqref{eq:layer_norm}.
The sequence of trainable layers forms a \gls{CNN}, which takes as input the matrix of bits $\Bm\in\{0,1\}^{N \times K}$, where $N$ corresponds to the 1D convolution dimension and $K$ to the number of input channels.
The last separable convolution has only two filters such that the output, of dimension $N\times 2$, can be converted to the baseband frequency symbols $\xv \in \CC^N$.

The neural receiver, depicted in Fig. \ref{fig:nn_rx_h}, uses a similar architecture.
The signal $\yv$ is first converted to a matrix of dimension $N \times 2$, which is processed by the \gls{CNN} as two different channels of dimension $N$.
The last separable convolution has $K$ filters, corresponding to the $K$ \glspl{LLR} that need to be predicted for every symbol. 
The exact parameters of each trainable component are listed in the next section.

%

\section{Evaluations} 
\label{sec:evaluations}

\subsection{Setup}
\label{sec:evaluation_setup}

\begin{table}

\centering

\newcolumntype{M}[1]{>{\centering\arraybackslash}m{#1}}
\newcolumntype{N}{@{}m{0pt}@{}}

\renewcommand{\arraystretch}{2}

\begin{tabular}{ |M{1.3cm}||M{0.7cm}|M{0.45cm}|M{0.45cm}|M{0.45cm}|M{0.45cm}|M{0.45cm}|M{0.7cm}| N }
 \hline
 	& Sep. Conv. & \multicolumn{5}{c|}{ResNet blocks} & Sep. Conv \\
 \hline
 Kernel size & 1 & 3 & 9 & 15 & 9 & 3 & 1\\
 \hline
 Dilation rate & 1 & 1 & 2 & 4 & 2 & 1& 1\\
 \hline
 \# Filters & \multicolumn{7}{c|}{128}\\
 \hline
\end{tabular}

\caption{Parameters for the \gls{CNN} transmitter and receiver.}
\label{table}
\renewcommand{\arraystretch}{1}

\end{table}

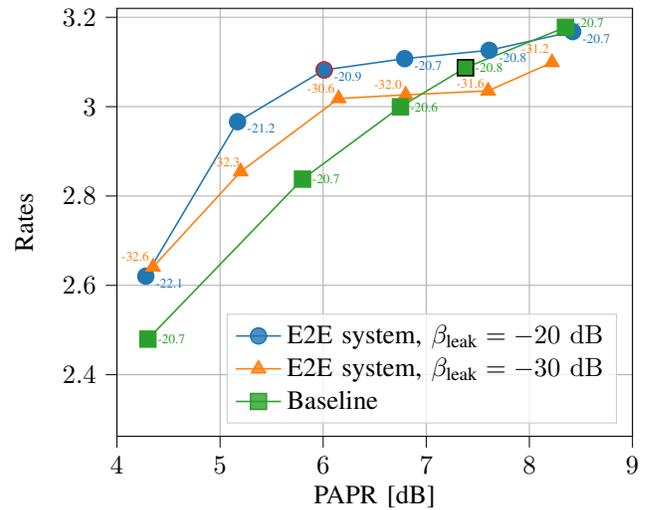
\begin{figure}[t]
\begin{tikzpicture}

\definecolor{color0}{rgb}{0.83921568627451,0.152941176470588,0.156862745098039}
\definecolor{color1}{rgb}{0.12156862745098,0.466666666666667,0.705882352941177}
\definecolor{color2}{rgb}{1,0.498039215686275,0.0549019607843137}
\definecolor{color3}{rgb}{0.172549019607843,0.627450980392157,0.172549019607843}

\begin{axis}[
legend cell align={left},
legend style={
  fill opacity=0.8,
  draw opacity=1,
  text opacity=1,
  at={(0.97,0.03)},
  anchor=south east,
  draw=white!80!black
},
tick align=outside,
tick pos=left,
x grid style={white!69.0196078431373!black},
xlabel={PAPR [dB]},
xmajorgrids,
xmin=4, xmax=9,
xtick style={color=black},
y grid style={white!69.0196078431373!black},
ylabel={Rates},
ymajorgrids,
ymin=2.2621444, ymax=3.22,
ytick style={color=black}
]
\addplot [semithick, color1, mark=*, mark size=3, mark options={solid}]
table {%
4.28 2.6205654
5.17 2.9663346
6.01 3.0821884
6.79 3.1074212
7.61 3.1261373
8.42 3.1679697
};
\addlegendentry{E2E system,  $\beta_{\text{leak}}=-20$ \si{dB}}
\draw (axis cs:4.28+0.05, 2.6205654-0.025) node[
  scale=0.5,
  anchor=base west,
  text=color1,
  rotate=0.0
]{-22.1};
\draw (axis cs:5.17+0.05, 2.9663346-0.025) node[
  scale=0.5,
  anchor=base west,
  text=color1,
  rotate=0.0
]{-21.2};
\draw (axis cs:6.01+0.05, 3.0821884-0.025) node[
  scale=0.5,
  anchor=base west,
  text=color1,
  rotate=0.0
]{-20.9};
\draw (axis cs:6.79+0.05, 3.1074212-0.025) node[
  scale=0.5,
  anchor=base west,
  text=color1,
  rotate=0.0
]{-20.7};
\draw (axis cs:7.61+0.05, 3.1261373-0.025) node[
  scale=0.5,
  anchor=base west,
  text=color1,
  rotate=0.0
]{-20.8};
\draw (axis cs:8.42+0.05, 3.1679697-0.025) node[
  scale=0.5,
  anchor=base west,
  text=color1,
  rotate=0.0
]{-20.7};
\draw (axis cs:4.35-0.35, 2.6411846+0.013) node[
  scale=0.5,
  anchor=base west,
  text=color2,
  rotate=0.0
]{-32.6};
\draw (axis cs:5.2-0.32, 2.8545365+0.013) node[
  scale=0.5,
  anchor=base west,
  text=color2,
  rotate=0.0
]{-32.3};
\draw (axis cs:6.15-0.35, 3.018253+0.013) node[
  scale=0.5,
  anchor=base west,
  text=color2,
  rotate=0.0
]{-30.6};
\draw (axis cs:6.8-0.35, 3.0264869+0.013) node[
  scale=0.5,
  anchor=base west,
  text=color2,
  rotate=0.0
]{-32.0};
\draw (axis cs:7.6-0.35, 3.0353181+0.006) node[
  scale=0.5,
  anchor=base west,
  text=color2,
  rotate=0.0
]{-31.6};
\draw (axis cs:8.2-0.32, 3.0989807+0.022) node[
  scale=0.5,
  anchor=base west,
  text=color2,
  rotate=0.0
]{-31.2};
\draw (axis cs:4.3+0.05 , 2.4796934-0.01) node[
  scale=0.5,
  anchor=base west,
  text=color3,
  rotate=0.0
]{-20.7};
\draw (axis cs:5.8+0.05 , 2.8377833-0.01) node[
  scale=0.5,
  anchor=base west,
  text=color3,
  rotate=0.0
]{-20.7};
\draw (axis cs:6.75+0.05 , 2.9996686-0.01) node[
  scale=0.5,
  anchor=base west,
  text=color3,
  rotate=0.0
]{-20.6};
\draw (axis cs:7.38+0.05 , 3.0871742-0.01) node[
  scale=0.5,
  anchor=base west,
  text=color3,
  rotate=0.0
]{-20.8};
\draw (axis cs:8.35+0.06 , 3.1775513) node[
  scale=0.5,
  anchor=base west,
  text=color3,
  rotate=0.0
]{-20.7};
\addplot [semithick, color2, mark=triangle*, mark size=3, mark options={solid}]
table {%
4.35 2.6411846
5.2 2.8545365
6.15 3.018253
6.8 3.0264869
7.6 3.0353181
8.22 3.0989807
};
\addlegendentry{E2E system,  $\beta_{\text{leak}}=-30$ \si{dB}}
\addplot [semithick, color3, mark=square*, mark size=3, mark options={solid}]
table {%
4.3  2.4796934
5.8  2.8377833
6.75  2.9996686
7.38  3.0871742
8.35  3.1775513
};
\addlegendentry{Baseline}
\addplot [draw=color0, draw=none, forget plot, mark=square]
table{%
x  y
-0.5 -0.5
0.5 -0.5
0.5 0.5
-0.5 0.5
-0.5 -0.5
};

\addplot [semithick, black, mark=square, mark size=3, mark options={solid,fill opacity=0}]
table {%
7.38 3.0871742
};

\addplot [semithick, color0, mark=o, mark size=3, mark options={solid,fill opacity=0}]
table {%
6.01 3.0821884
};

\end{axis}

\end{tikzpicture}
\caption{Rates achieved by the baseline and two E2E systems. The numbers represent the corresponding ACLRs.}
\label{fig:rates}
\vspace{-2pt}
\end{figure}

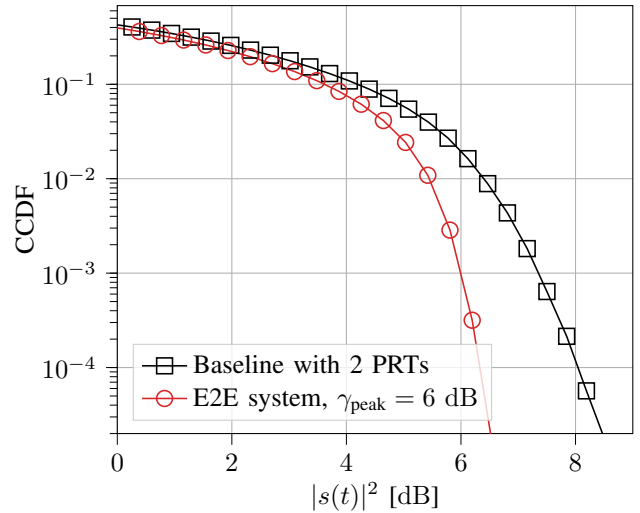
\begin{figure}[t]
\begin{tikzpicture}

\definecolor{color0}{rgb}{0.83921568627451,0.152941176470588,0.156862745098039}

\begin{axis}[
legend cell align={left},
legend style={
  fill opacity=0.8,
  draw opacity=1,
  text opacity=1,
  at={(0.03,0.03)},
  anchor=south west,
  draw=white!80!black
},log basis y={10},
tick align=outside,
tick pos=left,
x grid style={white!69.0196078431373!black},
xlabel={$|s(t)|^2$ [\si{dB}]},
xmajorgrids,
xmin=0, xmax=9,
xtick style={color=black},
y grid style={white!69.0196078431373!black},
ylabel={CCDF},
ymajorgrids,
ymin=2e-05, ymax=0.680933265155766,
ymode=log,
ytick style={color=black}
]
\addplot [semithick, black, mark=square, mark size=3, mark options={solid,fill opacity=0}]
table {%
-0.08820085 0.4336185714285714
0.25675178 0.4044919047619048
0.6017044 0.3749447619047619
0.94665706 0.3453609523809524
1.2916096 0.3158228571428571
1.6365623 0.28686047619047617
1.9815149 0.25822428571428574
2.3264675 0.2302204761904762
2.67142 0.20330619047619047
3.016373 0.17744904761904762
3.3613255 0.1528657142857143
3.706278 0.1298561904761905
4.051231 0.10853238095238095
4.3961835 0.08890952380952381
4.741136 0.07093333333333333
5.0860887 0.054555238095238094
5.4310412 0.039754285714285714
5.775994 0.02687095238095238
6.1209464 0.01630190476190476
6.465899 0.008869047619047618
6.8108516 0.004346666666666667
7.1558046 0.001819047619047619
7.500757 0.0006385714285714286
7.84571 0.00021428571428571427
8.190662 5.6666666666666664e-05
8.535615 1.5714285714285715e-05
};
\addlegendentry{Baseline with 2 PRTs}

\addplot [semithick, color0, mark=o, mark size=3, mark options={solid,fill opacity=0}]
table {%
-0.0046317433 0.39614533333333335
0.3829095 0.3622365714285714
0.7704508 0.3281958095238095
1.157992 0.2941921904761905
1.5455333 0.26081866666666664
1.9330745 0.22781142857142858
2.3206158 0.19607276190476192
2.708157 0.16539066666666666
3.0956984 0.13631580952380953
3.4832397 0.10925009523809524
3.8707807 0.08410971428571429
4.2583222 0.061540761904761904
4.6458635 0.041320190476190476
5.0334044 0.024225714285714284
5.4209456 0.010878095238095239
5.808487 0.002855809523809524
6.196028 0.00031714285714285715
6.5835695 1.1238095238095239e-05
};
\addlegendentry{E2E system, $\gamma_{\text{peak}}=6$~\si{dB}}

\end{axis}

\end{tikzpicture}
\caption{CCDF of the signal power for two systems having similar rates.}
\label{fig:ccdf}
\vspace{-8pt}
\end{figure}

\begin{figure}[t]
\input{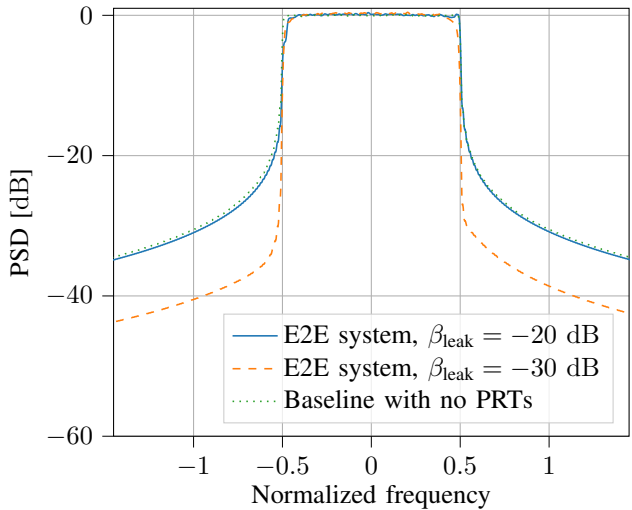}
\caption{PSD of three systems with similar PAPRs.}
\label{fig:psd}
\end{figure}

To evaluate the proposed scheme, an \gls{OFDM} system with $N=75$ subcarriers and an \gls{AWGN} channel is considered.
The \gls{SNR} of the transmission is set to 10~\si{dB}, both during training and evaluation.
The number of bits per channel use is set to $K=4$, and the baseline uses 16-QAM.

The transmitter and receiver share the same \gls{CNN} architecture, whose parameters are listed in Table \ref{table}.
It is worth noting that dilations are used to increase the receptive field of the \glspl{CNN}.
The optimization parameters are initialized at $\lambda_p^{(0)}=0, \lambda_l^{(0)}=0, \mu_p^{(0)}=0.1$, and $\mu_l^{(0)}=0.001$, and $\tau$ is set to $0.004$.
The difference in initialization between $\mu_p^{(0)}$ and $\mu_l^{(0)}$ reflects the fact that $L_{\beta_{\text{leak}}}(\thetav)$ is usually two orders of magnitude higher than  $L_{\gamma_{\text{peak}}}(\thetav)$.
The training process is composed of 2500 iterations, each including 15 \gls{SGD} steps are performed with a batch size of $B_s=1500$.
During training, $L_{\gamma_{\text{peak}}}(\thetav)$ is approximated using a temporal oversampling factor of $O_s=5$, which was found to be enough to accurately represent the continuous analog waveform \cite{922754}.
The baseline is evaluated for  $R \in \{0, 2, 4, 8, 16\}$ \glspl{PRT}, and the E2E system is trained for $\gamma_{\text{peak}} \in \{ 4, 5, 6, 7, 8, 9\}$~\si{dB} and two ACLR targets: $\beta_{\text{leak}}=-20$~\si{dB}, which corresponds to the baseline ACLR, and $\beta_{\text{leak}}=-30$~\si{dB}.

\subsection{Results}

In the following evaluations, the target \gls{PAPR} probability threshold in \eqref{eq:papr} is set to $\epsilon=10^{-3}$.
This is different from the value used for training, which was set to  $\epsilon=0$ in order to define a differentiable cost function.
Fig. \ref{fig:rates} shows the rates and \glspl{PAPR} achieved by the baseline and the two systems trained with target ACLRs of -20~\si{dB} and -30~\si{dB}.
The numbers next to the plots represent the \glspl{ACLR} of each evaluated system.
Firstly, it can be seen that all E2E systems satisfy the \gls{ACLR} targets. 
Secondly, for PAPRs lower than 7~\si{dB}, both trained systems achieve higher rates than the baseline, which is particularly interesting since the second system also achieves a 10~\si{dB} lower ACLR.
For PAPRs higher than 7.5~\si{dB}, the baseline and the system trained with $\beta_{\text{leak}} = -20$~\si{dB} enable similar performance.
Finally, it can be seen that the the training procedure allows to control the PAPR of the transmission.
Lower PAPR thresholds $\epsilon < 10^{-3}$ could be enforced by choosing a higher scalar $\tau$ during training. 

To visualize the effect of the \gls{PAPR} reduction on the signal energy, the \gls{CCDF} of two systems having similar rates are shown in Fig.\ref{fig:ccdf}.
These two systems are the baseline with 2 PRTs and the E2E system trained with $\gamma_{\text{peak}} = 6$~\si{dB} and $ \beta_{\text{leak}} = -20$~\si{dB},  respectively represented with a black square and a red circle in Fig. \ref{fig:rates}.
The \gls{CCDF} of the signal energy is expressed as
\begin{align}
\text{CCDF}_{|s(t)|^2}(x) = P \left( |s(t)|^2 > x \right).
\end{align}
It can be seen that the CCDF of the E2E system is significantly lower than the one of the baseline, which results in a difference of approximately 1.4~\si{dB} at a CCDF of $10^{-3}$.

Finally, the power spectral densities (PSDs) of three systems are shown in Fig.~\ref{fig:psd}.
The first one is the baseline with no \glspl{PRT}.
The second and third ones are E2E systems trained with the highest PAPR constraint ($\gamma_{\text{peak}}=9$~\si{dB}) and ACLR constraints respectively set to $\beta_{\text{leak}}=-20$~\si{dB} and $\beta_{\text{leak}}=-30$~\si{dB}.
It can be observed that the PSD of the baseline and that of the system trained with $\beta_{\text{leak}}=-20$~\si{dB} are nearly identical, but the PSD of the system trained with $\beta_{\text{leak}}=-30$~\si{dB} has considerably less out-of-band energy.

\section{Conclusion}
\label{sec:conclusion}

We have presented an E2E system that can be optimized to maximize an achievable information rate while satisfying constraints on the signal PAPR and ACLR.
The optimization objective and the two constraints were expressed as differentiable functions that can be minimized using \gls{SGD}.
The constrained optimization problem was solved using the augmented Lagrangian method, where the corresponding augmented Lagrangian is minimized iteratively.
Numerical results demonstrated the effectiveness of the proposed approach, with trained systems that meet ACLR and PAPR targets while achieving rates comparatively higher than the baseline.
This work can be seen as a first step towards learning of new waveforms for future generations of ACLR- and PAPR-constrained systems.
Scaling to more subcarriers and the integration of more realistic channels constitute possible research directions.


\bibliographystyle{IEEEtran}
\bibliography{IEEEabrv, bib_abrv, bibliography}

\end{NoHyper}
\end{document}